\documentclass[aip,pop,reprint]{revtex4-2}

\newcommand{\PID}{\ensuremath{\Pi D}}
\newcommand{\ptheta}{{\ensuremath{p \theta}}}
\newcommand{\divq}{\ensuremath{\partial_m q_{xxm}}}
\newcommand{\divqi}{\ensuremath{\partial_m q_{i,xxm}}}

\usepackage{amsmath,amssymb}
\usepackage{graphicx} 

\usepackage[normalem]{ulem}

\usepackage{orcidlink}

\DeclareUnicodeCharacter{2212}{\ensuremath{-}}

\usepackage{url} 

\usepackage{hyperref}
\usepackage{todonotes}
\hypersetup{
     colorlinks = true,
     pdfborder={0 0 0},
     bookmarksopen=true,
     linktocpage=true,
     linkcolor=blue,
     citecolor=blue,
     urlcolor=blue,
}

\usepackage[dvipsnames]{xcolor}
\definecolor{OliveGreen}{rgb}{0,0.6,0}

\begin{document}

\title{The effects of closure parameter on the properties of kinetic range plasma turbulence}

\author{Jago V. H. Edyvean \orcidlink{0009-0003-6942-5410} }
\email[Corresponding author: ]{jagoedyvean1@gmail.com}
\affiliation{Victoria University of Wellington, Kelburn, 6021, NZ}

\author{Andrea F. Irwin \orcidlink{0009-0001-7900-0593} }
\affiliation{Victoria University of Wellington, Kelburn, 6021, NZ}

\author{Tulasi N. Parashar \orcidlink{0000-0003-0602-8381} }
\affiliation{Victoria University of Wellington, Kelburn, 6021, NZ}

\begin{abstract}

Simulations of plasmas such as the solar wind require an accurate representation of kinetic physics given the lack of frequent collisions. Fully kinetic treatments are computationally expensive and necessarily confine simulations to small domain sizes. A good compromise is to utilise higher-order fluid-moment models such as the 10-moment model.

This model retains all physics up to the second moment, with the moment hierarchy closed by approximating the heat flux.

One choice is to use a model for the divergence of the heat flux, $\partial_m q_{ijm} \approx v_t |k_0|(\mathbf{P}_{ij}-p\delta_{ij})$, with a free parameter $k_0$, called the ``closure parameter''.
We study the effects of this closure parameter on the properties of turbulence, combined with artificial ion-electron mass ratios. We show that the heating of the two plasma species is affected by the choice of the closure parameter. For all combinations of closure parameters, the relative compressive and incompressive heating rates are affected even for realistic mass ratios. The effects are more prominent for lower mass ratios and diminish when approaching realistic mass ratios. 
\end{abstract}

\maketitle


\section{Introduction}\label{Ch:Intro}

Plasmas are found throughout the observable universe, from the solar wind that surrounds Earth \citep{parker1958APJ} to the intracluster medium \citep{lyutikov2007APJ}. The mean free paths for many of these systems are extremely large, rendering the role of collisions insignificant \citep{hollweg1971PRL,marsch2006LRSP}. Most of these systems are either observed, or believed, to be in a turbulent state \citep{coleman1968turbulence, galtier2018turbulence,matthaeusPoP21}. The solar wind, being close enough to Earth for in-situ measurements, presents as an ideal natural plasma laboratory to investigate kinetic turbulent space plasmas \cite{brunocarboneLRSP13}.

A long-standing problem in kinetic plasma physics is the heating of such plasmas. How are the turbulent fluctuations converted into internal degrees of freedom without significant collisions \citep{matthaeusPoP21}? Answering this question has major implications for the coronal heating \citep{parker1958APJ, parker1960ApJ} and solar wind heating problems \citep{freemanGRL1988}. The question entails many details, such as the nature of processes responsible for this conversion, the nature of the turbulent cascade below ion kinetic scales, and the partitioning of the dissipated energy between ions and electrons \citep{kiyani2015RS, matthaeusPoP21}. In-situ observations \citep[e.g.][ and references therein]{osmanPRL2011, kiyani2015RS, MANZINI24PRL} as well as simulations \citep[e.g.][and references therein]{parasharPoP2009, tenbargeAPJL2013, karimabadi13PoP, told2015multiscale} have brought forth a great deal of understanding but disagreements persist \citep{matthaeusPoP21}. 

Fully-kinetic simulations of multi-species turbulence are computationally expensive. This limits kinetic models to either small domain sizes, coarse and under-resolved grids, and other artificial simulation parameters. A common compromise is to use artificially heavy electrons and an artificial speed of light \citep{gary2016ApJ,verscharen2020MNRAS}. Other approaches to reduce the computational expense while approximately keeping the relevant physics in the model include multi-fluid MHD models \citep{liu93GRL}, multi-moment fluid models \citep{dong19GRL} and hybrid kinetic models \citep{winske2003hybrid} to name a few approximations. 

The moment hierarchy of the Vlasov equation, combined with Maxwell's equations, represents exact but incomplete physics of the system \citep{freidberg1982RMP}, requiring a model for the (n+1)th moment if n moments are kept in the hierarchy. Hence, this incompleteness depends on the number of moments taken; fluid models with more moments retain more of the kinetic physics \citep{grad1949kinetic, schunk1977RoG, barakat1982PP, hakim2006JCP, hakim2008JFE}. A popular choice for these higher-order fluid-moment models is to keep the first three moments (0,1,2) up to the pressure tensor, along with a model for the heat-flux e.g. to represent the physics of Landau damping (hence the name Landau-Fluid models) \citep{hammettPRL90, snyder1997PoP, passot2004PoP, goswami2005PoP, hakim2008JFE}.

Taking the first three moments of the Vlasov equation gives us the following equations \citep{hakim2008JFE} for a given species (species labels omitted),
\begin{eqnarray}
     \frac{\partial n}{\partial t} + \frac{\partial}{\partial x_j}(nu_j)& = & 0 \\
     \label{m0}
     m \frac{\partial}{\partial t}(nu_i) + \frac{\partial \mathcal{P}_{ij}}{\partial x_j} & = & nq(E_j + \epsilon_{ijk} u_j B_k)  \\
     \label{m1}
    \frac{\partial \mathcal{P}_{ij}}{\partial t} + \frac{\partial \mathcal{Q}_{ijk}}{\partial x_k}& = & nqu_{[i}E_{j]} + \frac{q}{m} \epsilon_{[ikl}\mathcal{P}_{kj]}B_l
    \label{m2}
\end{eqnarray}
where $q$ is charge, $m$ is mass, $n$ is number density, $u$ is flow velocity, $E$ and $B$ are the electric and magnetic field respectively, $\mathcal{P}_{ij} \equiv m\int v_i v_j f d^3 v  $ is the second moment of the distribution function, $\mathcal{Q}_{ijk} \equiv m \int v_i v_j v_k f d^3 v $ is the third moment of the distribution function. Repeated indices are summed over and the square brackets denote a sum over permutations of the indices. The third moment tensor can be written $\mathcal{Q}_{ijk} = q_{ijk} + u_{[i} \mathcal{P}_{jk]} - 2mnu_iu_ju_k$, where $q_{ijk} \equiv m \int (v_i-u_i)(v_j-u_j)(v_k-u_k)fd^3v$ is the heat flux tensor. These fluid equations are coupled with Maxwell's equations.

One way to address the closure problem is to set the (n+1)th moment to zero, called truncation. This model was proposed in \textcite{grad1949kinetic} for a 13-moment spectral method, using exponential Gaussian weighted Hermite functions in velocity space together with a truncation closure (i.e the 14th moment is set to zero). Truncation proves to be remarkably effective as a method for closing certain Hermite moment models, while failing to simultaneously conserve mass, momentum, and energy in others \citep{issan2024conservative}. Many modern multi-moment models keep moments only up to the pressure tensor, and approximate the heat flux. Various sensible approximations \citep{hammettPRL90, goswami2005PoP, hakim2008JFE, wang2015comparison, ng2020improved, issan2024conservative} have been tried to achieve good conservation properties.

\begin{table*}[!ht]
\centering
\begin{tabular}{|c |c |c |c |c |c |c |c |c |c |c |c |}
 \hline
 $d_e$ & $n_0$ & $c$ &  $\omega_{pe}$ & $v_{Ai}/c$ & $eps$ & $dt/\omega_{pe}$ & $\Delta B/B_0$ & $\beta_i$ & $\beta_e$ &$L/d_i$ &  $N_x, N_y$\\ 
 \hline
  1 & 1 & 1 & 1 & 0.0115 & 0.2 & 0.2 & 0.2 & 0.1 & 0.1   &  8$\pi$&2048\\ 
 \hline 
\end{tabular}
\caption{Normalisations used and parameters kept constant across all 10-moment two-fluid Gkyell simulations.}
\label{tab:paramconst}
\end{table*}
\begin{table*}[ht!]
\centering
\begin{tabular}{|c |c |c |c |c |c |c |c |c |c |c |c |c |c |}
 \hline
 $\boldsymbol{m_{i} / m_{e}}$ & $\boldsymbol{v_{Ae}}$ &  $\boldsymbol{B_{0}}$  & $\boldsymbol{\Delta B}$  & $\boldsymbol{\omega_{ci}}$ & $\boldsymbol{\omega_{ce}}$ & $\boldsymbol{d_{i}}$  & $\boldsymbol{L}$ & $\boldsymbol{\tau_{0}}$ & $\boldsymbol{dx}$  \\ 
 \hline
 25   & 5.74$\times 10^{-2}$ & 5.74$\times 10^{-2}$  & 1.15$\times 10^{-2}$ & 2.30$\times 10^{-3}$  & 5.74$\times 10^{-2}$ & 5    &  125.66  & 8.7$\times 10^{3}$ &  6.14$\times 10^{-2}$  \\ 
 \hline
 250  & 1.81$\times 10^{-1}$ & 1.81$\times 10^{-1}$ & 3.64$\times 10^{-2}$ & 7.27$\times 10^{-4}$ & 1.81 $\times 10^{-1}$ & 15.8 &  397.38  &  2.75$\times 10^{4}$ & 1.94 $\times 10^{-1}$ \\
 \hline
 1000 & 3.64$\times 10^{-1}$ & 3.64$\times 10^{-1}$ & 7.27$\times 10^{-2}$ & 3.64$\times 10^{-4}$ & 3.64$\times 10^{-1}$ & 31.6 &  794.77  &  5.50$\times 10^{4}$ & 3.88$\times 10^{-1}$  \\
 \hline 
 1836 & 4.93$\times 10^{-1}$ & 4.93$\times 10^{-1}$ & 9.86$\times 10^{-2}$ & 2.68$\times 10^{-4}$ & 4.93$\times 10^{-1}$ & 42.8  &  1076.90  &  7.45$\times 10^{4}$ & 5.26$\times 10^{-1}$ \\ 
 \hline 
\end{tabular}
\caption{Parameters for the 10-moment two-fluid Gkyell simulations: $m_i/m_e$ is the mass ratio,the electron Alfv\'en velocity is $v_{Ae} [c]$, the background field strength is $B_{0} [c$ $m_{e}^{1/2}\mu_0^{1/2}n_0^{1/2}]$, magnetic field fluctuations is $\Delta B [c$ $m_{e}^{1/2}\mu_0^{1/2}n_0^{1/2}]$, ion cyclotron frequency is $\omega_{ci} [\omega_{pe}]$, electron cyclotron frequency is $\omega_{ce} [\omega_{pe}]$, ion inertial length is $d_i [d_e]$, domain length is $L [d_e]$, eddy turnover time is $\tau_0 [d_e c^{-1}]$ and grid size is $dx [d_e]$.}
\label{tab:param}
\end{table*}
\begin{center}
\begin{table}[ht!]
\begin{tabular}{||c |c |c ||}
 \hline
 $m_i/m_e$ & Ions, $\boldsymbol{k_{0i}}$ & Electrons, $\boldsymbol{k_{0e}}$\\ 
 \hline
  25, 250, 1000, 1836 & $1/d_e$ & $1/d_e$ \\
  \hline
  25, 250, 1000, 1836 & $1/d_e$ & $1/d_i$ \\
  \hline
  25, 250, 1000, 1836 & $1/d_i$ & $1/d_e$ \\
  \hline
  25, 250, 1000, 1836 & $1/d_i$ & $1/d_i$ \\
 \hline 
\end{tabular}
\caption{Combinations of mass ratios and closure parameters}
\label{tab:comb}
\end{table}
\end{center}  

Most of the works dealing with closure of the heat-flux attempt to model the linear solutions of the Vlasov equation, trying to capture the physics of linear Landau damping as accurately as possible \citep[e.g.][and references therein]{hammettPRL90, goswami2005PoP, hunana2018newclosures}. For collisionless space plasma systems, closures which have their basis in small mean-free-path approximation, such as the Chapman-Enskog expansion \citep{kersonhuangbook}, are unsuitable. In order to broach the problem of collisionless closures, one can develop closures that give the (exact) kinetic results for linear problems. There are two main routes taken to this effect \citep{wang2015comparison}. The first of these, \citep{chang92PoFPP}, looks to find the exact solution to the linearised Vlasov-Maxwell equations. This leads to expressions which are complicated functions of the plasma response function, $R(\zeta)=1+\zeta Z(\zeta)$, where $Z(\zeta)=\pi^{-1}\int dt\,exp(-t^2)/(t-\zeta)$ is the plasma dispersion function in Fourier space, and their physical space expressions being equivalent to a non-local closure involving integration along field lines or across all space. The second route, \citep{hammettPRL90}, uses an approximation of the solution to the Vlasov-Maxwell equations based on a $n$-pole Pade-series representation of the response function. The Hammett-Perkins type closure also leads to non-local closures, requiring integration along field lines, and thus further, local approximations, must be implemented.

Using a local, linear approximation to the Hammett-Perkins closure, \textcite{wang2015comparison} came up with a much simplified representation of the divergence of heat flux
\begin{equation}
    \partial_m q_{ijm} \approx v_t |k_0|(\mathbf{P}_{ij}-p\delta_{ij}).
    \label{equ:divhf}
\end{equation}
Given the dependence on thermal speed \textcite{wang2015comparison} described $v_t |k_0|$ as “... superficially, like collisional relaxations of the pressure tensor...”. However, literally following the definition in Eqn. \ref{equ:divhf}, we prefer to think of $1/k_{0\alpha}$ as a scale over which the anisotropic internal energy imbalances, captured by ($P_{ij} – p\delta_{ij}$), are transported by thermal motions \textcite{wang2015comparison} showed that $k_{0i}=k_{0e}=1/d_e$ choice for closure parameters in the 10-moment model describes reconnection physics consistent with PIC simulations, motivating the use of these values in further 10-moment models \cite{ng2017POP, dong19GRL, edyvean24scale}. It is worth noting that \textcite{ng2020improved} suggested further improvements of the closure model by incorporating the physics of temperature-gradient driven dynamics. However, in this study we continue using the Wang closure model to perform direct comparison of our conclusions. However, in this study we continue using the \textcite{wang2015comparison} closure model to identify how our previous conclusions are affected by the choice of
$k_0$ \citep{edyvean24scale}. Even though the $1/d_e$ choice for both species was suggested to be optimal, the effects of varying $k_0$ on turbulence properties have not be quantified. Hence we ask the questions: How do various choices of $k_0$ affect the dynamics of kinetic range plasma turbulence and heating of ions and electrons? Is there a mass ratio beyond which different choices of $k_0$ converge to similar behaviours?

\section{Simulation Setup}\label{simulations}

\begin{figure*}[ht!]
    \centering
    \includegraphics{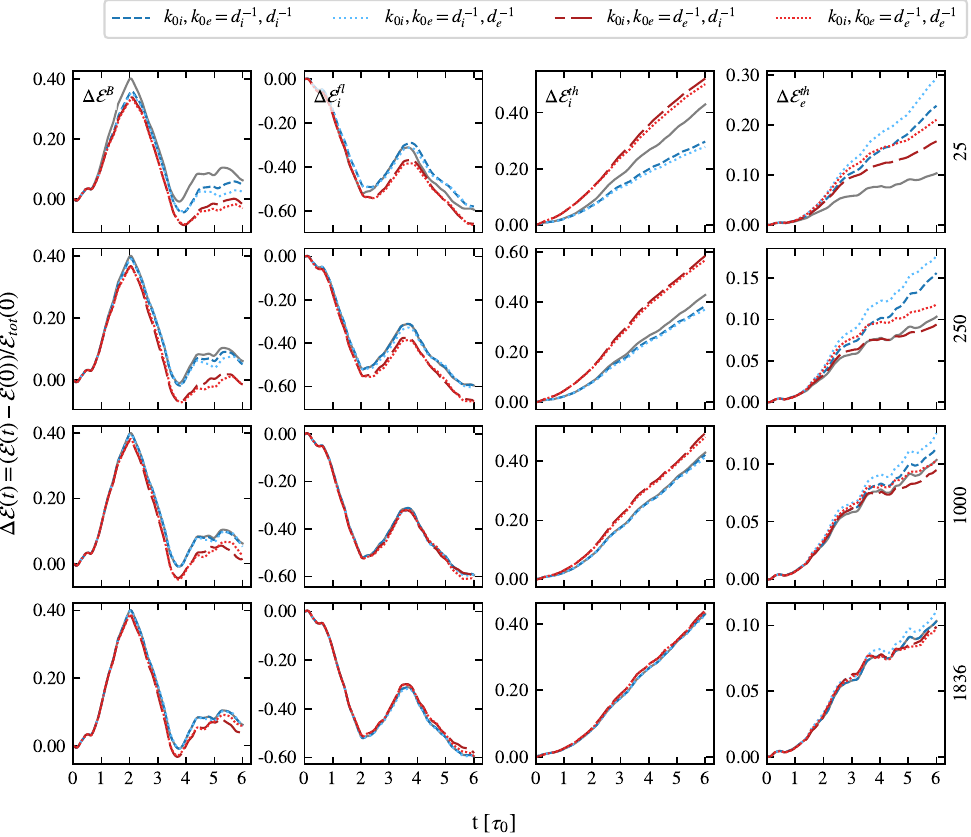}
    \caption{Time evolution of the magnetic field fluctuation energy, $\Delta \mathcal{E}^{B}$, the ion flow energy, $\Delta \mathcal{E}_{i}^{fl}$, the ion thermal energy, $\Delta \mathcal{E}_{i}^{th}$, and the electron thermal energy, $\Delta \mathcal{E}_{e}^{th}$, for different $m_i/m_e$ and combinations of $k_{0i,0e}$. All values are normalised to $\mathcal{E}_{tot}(0)=\mathcal{E}^{B}(0) + \mathcal{E}^{fl}_{i}(0) + \mathcal{E}^{fl}_{e}(0)$.
    The solid gray line plots the values for $k_{0i}, k_{0e} = d_i^{-1}, d_i^{-1}$, $m_i/m_e = 1836$, included as a reference to show convergence.}
    \label{fig:big_energies}
\end{figure*}

We use the Gkeyll 10-moment model for the simulations performed in this study. The model has used to study various problems such as island coalescence \citep{ngPoP2017}, kinetic drift instabilities \citep{ng2019JGRSP}, and turbulence in 3D asymmetric reconnection \citep{tenbarge2019JGRSP}. \textcite{tenbarge2019JGRSP} compared the 10-moment model with kinetic particle-in-cell simulations and MMS data and found that the 10-moment model compares favorably with the kinetic model and data. These kinetic validations make this model an enticing choice for the study of kinetic range dynamics while reducing the required computational load. 

The simulations performed in this study are in the 2.5D geometry, keeping three components for all vector quantities, e.g. $\boldsymbol{u}=(u_x, u_y, u_z)$, while retaining two spatial, $\boldsymbol{r}=(x, y)$, dimensions. The simulations are initialised using the Orszag-Tang vortex \citep{orszagJFM79} initial conditions.

\begin{eqnarray*}
    u_x(x,y,0) &=& -\Delta u \sin(2\pi y/L) \\
    u_y(x,y,0) &=& \Delta u \sin(2\pi x/L) \\
    B_x(x,y,0) &=& -\Delta B \sin(2\pi y/L) \\
    B_y(x,y,0) &=& \Delta B \sin(4\pi x/L)
\end{eqnarray*}

These initial conditions allow rapid development of turbulence and maximise the available bandwidth, especially for simulations with limited inertial range \citep{parasharPoP2009,parasharApJ15,li2019JPP}. The simulations are evolved up to 6$\tau_0$. The peak of the mean squared current, which in magnetohydrodynamics would measure the peak of dissipation, happens at about 2$\tau_0$. After 4$\tau_0$ the simulations have fully developed spectra and faithfully capture turbulent decay dynamics. Hence, we perform all our analyses after $4\tau_0$, where $\tau_0=L_0/\Delta u = L/(2\pi \Delta u)$ is the eddy turnover time where $\Delta u$ is the velocity fluctuation strength and $L$ is the system size.

The normalisations as well as the parameters kept constant across all simulations are given in the table \ref{tab:paramconst}, where $\omega_{pe}$ is the electron plasma frequency, $v_{Ai}$ is the ion Alfv\'en speed, $dt$ is the time step, $eps$ is the turbulent fluctuation strength, $\beta$ is the plasma beta, $d_e$ is the electron inertial length, $n_0$ is the number density, $c$ is the speed of light, $\mu_0$ is the permeability of free space, $\epsilon_0$ is the permittivity of free space and $\beta_i/\beta_e$ is the ratio of ion and electron plasma betas. 

Table \ref{tab:param} shows the parameters that change as the mass ratio is changed and table \ref{tab:comb} shows the different combinations of mass ratios and closure parameters studied. As we wish to perform controlled numerical experiments, while having the same inertial-range dynamics across simulations, we fix the system size in terms of the ion inertial length, $\beta_i$, $\beta_e$, and keep a constant $v_{Ai}/c$. Keeping $\beta_i$ and $\beta_e$ fixed then requires fixing $v_{ti}$, $v_{te}$, hence we vary $B_0$ to ensure all the appropriate parameters are the same across the simulations. For further details of all parameters kept fixed and the rationale behind those please see the precursor to this study \citep{edyvean24scale}. All-in-all four different species specific closure parameter combined with four different choices of $m_i/m_e$ give us a set of 16 simulations.

In order to perform controlled numerical experiments that retain the same inertial range dynamics, while keeping the numerical parameters the same, we choose to fix the system size in terms of ion inertial length ($d_i=c/\omega_{pi}$). We also have exactly the same number of grid points across all simulations, giving us the same grid spacing in units of $d_i$. Therefore, the number of grid points is theoretically dictated by the realistic mass ratio case and the electron scales are larger and significantly over-resolved for the smallest mass ratio case. This, with each simulation having a domain length of $L=8\pi d_i$ and 2048$^2$ grid points, allows the same amount of the inertial range to be captured. To enable comparisons with other works, we fix the ratio of the ion Alfv\'en speed to the speed of light while the electron Alfv\'en speed changes with mass ratio. $V_{Ae}$ becomes roughly half the speed of light in the realistic mass-ratio case. Ideally, this requires a relativistic treatment, but that is beyond the scope of this work. 

Although one would ideally use a fully kinetic simulation with realistic mass ratio as the ground truth for such comparisons, the stringent requirement of achieving realist mass ratio while keeping many parameters fixed meant that we could not use our existing fully kinetic simulations with lower mass ratio in this study or perform a new fully kinetic simulation. For realistic mass ratio, a comparison of fully kinetic simulations to hybrid-kinetic simulations with an electron closure model has only been done in the linear regime of Landau damping \citep{lautenbachJPP26}. In the non-linear regime all the comparisons have been at reduced mass ratios \citep[for example]{wang2015comparison, ng2020improved, Lautenbach}. We are in the process of designing a set of 10-moment and fully kinetic simulations as a follow-up to this study. Hence, in this study, we use the mr=1836 10-moment simulation as the canonical simulation, even though it does not necessarily contain the same physics as the fully kinetic model, and all comparisons will be made in reference to this.

\section{Results}\label{results}

Figure \ref{fig:big_energies} shows the change in various energies (magnetic $\Delta\mathcal{E}_B$, ion flow $\Delta \mathcal{E}_{i}^{fl}$, ion thermal $\Delta \mathcal{E}_{i}^{th}$, and electron thermal $\Delta \mathcal{E}_{e}^{th}$) as a fraction of the initially available fluctuation energy ($\mathcal{E}_{tot}(0) = \mathcal{E}_B(0)+\mathcal{E}_i^{fl}(0)+\mathcal{E}_e^{th}(0)$). We do not show the electron flow energy because it is a negligible part of the total energy budget. The rows show these quantities for various mass ratios and the columns show various combinations of $k_{0i},k_{0e}$. The blue curves show cases where the ions have $k_{0i}=d_i^{-1}$ and the red curves show cases where $k_{0i}=d_e^{-1}$. Similarly, the dashed curves show where $k_{0e}=d_i^{-1}$ and the dotted curves show where $k_{0e}=d_e^{-1}$.

\textcite{wang2015comparison} suggested $k_{0i},k_{0e}=d_e^{-1},d_e^{-1}$ as a good choice. Naively, one would anticipate $k_{0i},k_{0e}=d_i^{-1},d_e^{-1}$ as the natural choice. However, if we anticipate the scale of heat-flux divergence to be comparable to the scales at which significant dissipation happens in plasmas, fully kinetic particle-in-cell simulations as well as MMS observations make preliminary suggestions that $k_{0i},k_{0e}=d_i^{-1},d_i^{-1}$ might be the most appropriate choice \citep{matthaeusApJ2020,yangApJ2022,MANZINI24PRL}. In order to determine the closure with the most accurate dynamics, analysis against comparable fully kinetic simulations must be performed. While the simulations performed in this study have not been compared against equivalent fully kinetic mr=1836 simulations, the analysis suggests that for a realistic mass ratio, using $k_{0i}, k_{0e} = d_i^{-1}$ is a good choice for our “ground truth” simulation. One could naively argue the scale of heat-flux divergence to be comparable to the scales of heating. Based on this, one could argue that $k_{0i}, k_{0e} = d_i^{−1}, d_i^{-1}$ might be a sensible choice for the closure parameter. Hence, the gray curve in each panel shows the corresponding quantity for the $m_i/m_e=1836$ case with $k_{0i},k_{0e}=d_i^{-1},d_i^{-1}$. 

The variation of the closure parameter seems to have little effect on larger mass ratios where the ion and electron scales are well separated. However, for smaller mass ratios, all energies show different evolutions, with the differences being more prominent for the lowest mass ratio. Particularly, the kinetic energy after $t\sim 2\tau_0$, after the peak of the mean squared current, and the magnetic energy after $t\sim 4\tau_0$, in the fully developed regime, show significant variability depending on which combination of closure parameters is used. Thermal energies\footnote{We use ``thermal energy'' and ``internal energy'' interchangeably in this manuscript.} show significant differences much earlier. 

All combinations of closure parameters for $m_i/m_e=25$ show major differences with the 1836 case. However, interestingly, the $k_{0i}=d_i^{-1}$ cases for $m_i/m_e=250$ converge quickly towards the 1836 results, except for the electron internal energy. The $k_{0i}=d_e^{-1}$ cases continue showing minor differences in magnetic and ion kinetic energies, even though negligible, even for larger mass ratios. For cases where $k_{0i}=d_e^{-1}$, the ion heat-flux divergence is confined to electron scales. As the mass ratio increases and electron scales become progressively smaller, the scale separation between ions and electrons in turn limits the dispersion-scales of ion internal energy through the heat-flux channel. As the energy dissipated into the ions saturates to a consistent value at mr=256 and above, the electrons see a consistent cascade of the remaining energy and hence their heating is also consistent. Changing $k_{0i}$ to $d_i^{-1}$  implies that the dispersion of energy through the heat-flux channel happens at ion scales, significantly affecting ion-heating channels. Hence, a mass-ratio convergence to the canonical simulation is no longer observed. The error in electron internal energies, even though significant in a relative sense, is about 5\% in terms of the overall fluctuation energy. This could potentially happen for various reasons including but not limited to: i) the electron kinetic scales being barely resolved in the 1836 case ($\delta x \sim 0.56 d_e$), or ii) the lack of relativistic effects affecting the large electron thermal speeds, or iii) the limitations of the closure model itself.

\begin{center} 
\begin{figure}[t!]
    \includegraphics[width=3.5in]{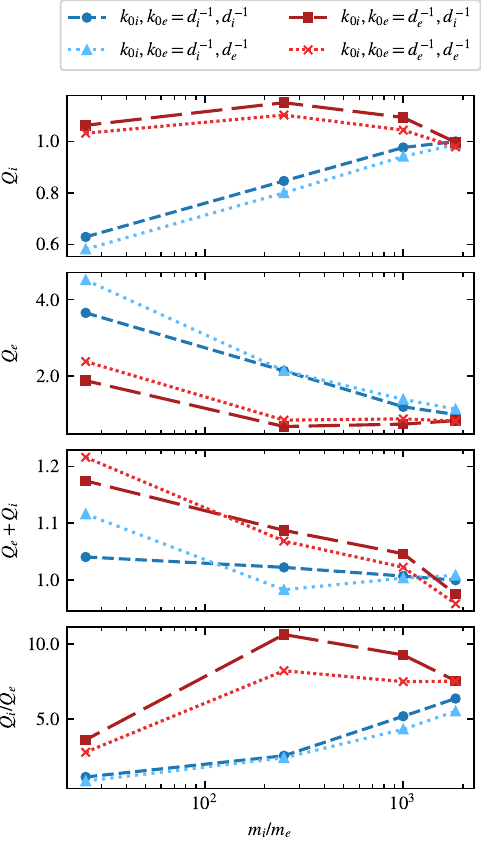}
    \caption{The mean average ion and electron heating rates for increasing $m_i/m_e$ and different combinations of $k_{0e,0i}$, between 4$\tau_0$ and 6$\tau_0$. The top panel shows the average electron heating rate, $\langle Q_e \rangle$, the second to top shows the average ion heating rate, $\langle Q_i \rangle$, the third from top shows the sum of the average ion and average electron heating, $\langle Q_i \rangle + \langle Q_e \rangle$, and the bottom panel shows the ratio of average ion heating to average electron heating, $\langle Q_i \rangle / \langle Q_e \rangle$. Different closure parameter combinations are shown by different symbols and linestyles, defined in the legend.}
    \label{fig:heating}
\end{figure}
\end{center}

\begin{figure*}
    \centering
    \includegraphics{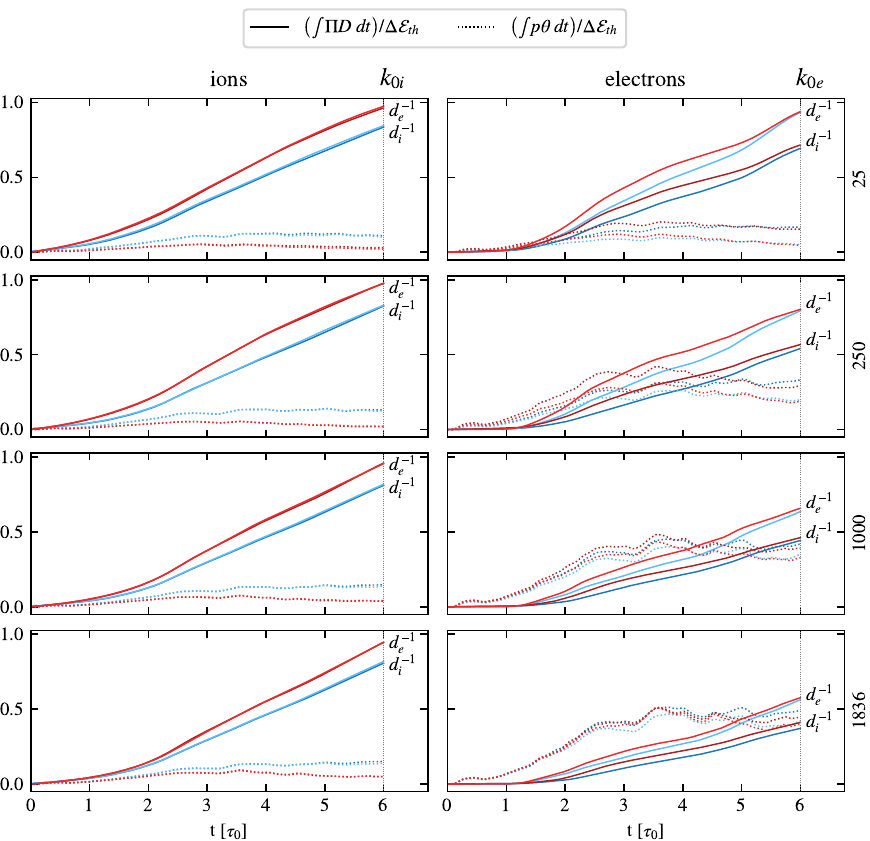}
    \caption{The cumulative $\Pi D$ (represented with a solid line) and $p\theta$ (represented with a dashed line) for different $m_i/m_e$ and combinations of $k_{0e,0i}$, for both ions and electrons. All values are normalised to the total change in the thermal energy of the appropriate species for each simulation.} 
    \label{fig:PiDpTh}
\end{figure*}

Figure \ref{fig:heating} shows the heating rates for all the simulations performed. The heating rates are computed as
$$
Q_\alpha = \bigg\langle\frac{d \mathcal{E}_\alpha^{th}}{dt}\bigg\rangle_{4\tau_0\le t\le6\tau_0}
$$
While plotting, we normalise $Q_i$, $Q_e$, and $Q_i+Q_e$ by the corresponding values for the $m_i/m_e=1836$, $k_{0i},k_{0e}=d_i^{-1},d_i^{-1}$ simulation to show convergences towards our anticipated appropriate closure parameter choice. $Q_i/Q_e$ values are naturally dimensionless and not further normalised. The closure parameter variation has the most significant effect on the heating rates for the lower mass ratios, with a significant scatter in all heating rates ($Q_i$,$Q_e$,$Q_i+Q_e$). The largest variation is seen in electron heating. For the lowest mass ratio, the ion heating can vary by as much as 40\%, while the electron heating can vary by more than 100\%. The total heating rate shows a better trend with the worst error being $\sim$ 20\%. The scatter reduces with increasing mass ratio, with the heating rates almost converging for all closure parameter combinations at $m_i/m_e=1836$. The relative heating of ions and electrons does not seem to converge even for realistic mass ratios. Interestingly, the largest scatter for $Q_i/Q_e$ happens at $m_i/m_e=250$. 

Changing the ion closure parameter $k_{0i}$ between $d_i^{-1}$ and $d_e^{-1}$ has the most drastic effect on the heating rates (see the trends of blue curves versus red curves). Changing the electron closure parameter $k_{0e}$ for fixed $k_{0i}$, on the other hand, does not seem to affect the heating rates significantly (see the trends of dotted versus dashed curves for the same colour). This is perhaps expected, given that the ions see the cascaded energy first in the scale-space. It is fair to conclude, in light of recent studies \citep{yangApJ2022,MANZINI24PRL} which show that the ions as well as electrons start heating near $d_i$, that the ion closure parameter is more important than the electron closure parameter, with $k_{0i}=d_i^{-1}$ being a sensible choice. Interestingly, the $k_{0i},k_{0e}=d_i^{-1},d_i^{-1}$ choice shows consistent total heating across all mass ratios. This trend also appears to support the suggestions made by the cited studies regarding both species starting to heat at the ion scales.

Having analysed the overall heating properties of ions and electrons, we proceed to investigate the relative contributions of the compressive and viscous-like processes to heating \citep{yangApJ2022}. The pressure-strain interaction $\mathbf{P}\cdot\mathbf{\nabla}\cdot\mathbf{u}$ responsible for converting energy from bulk flow motions into internal degrees of freedom can be decomposed into a compressive ($p\theta = Tr(\mathbf{P})Tr(\mathbf{S})$) and a viscous-like part, $\Pi D = \Pi_{ij}D_{ij}$, where $\mathbf{P}$ is the pressure tensor, $\mathbf{S}$ is the velocity strain tensor, $\Pi_{ij}$ is the traceless part of the pressure tensor, and $D_{ij}$ is the symmetric part of the strain tensor (see e.g. \cite{matthaeusApJ2020} and the references in there for more details). 

Figure \ref{fig:PiDpTh} shows the time-integrated values of \PID~ (solid lines) and \ptheta~ (dotted lines) normalized to the change in corresponding $\mathcal{E}_{th}$ for all simulations. The left column shows ions, and the right column shows electrons. Each row corresponds to a mass ratio identified on the right of each row. The colorscheme is the same as all the other figures, but for convenience we label the relevant curves with the value of the closure parameter for ions (electrons) in the corresponding column. Normalizing the values to the change in internal energy allows us to compare the relative contributions of \PID~ and \ptheta~ to heating for each simulation. The time integrated \ptheta~ saturates for both ions and electrons at $t\sim2.5\tau_0$ for all combinations of mass ratios and closure parameters. This implies that the dissipative processes for $t>2.5\tau_0$ are primarily \PID~ driven and the system is likely in a state of near-incompressibility \citep{matthaeus1988nearly}.

The contribution of \PID~ for ions is significantly large compared to \ptheta~ for all the mass ratios. Interestingly, the mass ratio does not appear to affect this split significantly. However, changing $k_{0i}$ from $d_i^{-1}$ to $d_e^{-1}$ increases the relative contribution of the incompressive (\PID) heating fraction. As $k_0i$ becomes comparable to $d_i^{-1}$, the transport of heat by the thermal motions as represented by the $\partial_mq_{ijm}$ term can potentially enhance the dilatation mechanisms and proportionally suppressing the viscous-like contribution. The electrons, on the other hand, show a much more complex behaviour across mass ratios. \PID~ dominates \ptheta~ for small mass ratios while \ptheta~ dominates for larger mass ratios. See \cite{edyvean24scale} for a more detailed discussion of the effects of varying mass ratio on heating. Changing the electron closure parameter ($k_{0e}$) from $d_i^{-1}$ to $d_e^{-1}$ increases the relative contribution of \PID~ similar to ions. Also similar to the ions, changing the closure parameter for the other species ($k_{0i}$ in this context) does not significantly affect the relative contributions of \PID~ and \ptheta. Interestingly, the late time surpassing of \PID~ over \ptheta~ for large mass ratios found in \cite{edyvean24scale} is true only for $k_{0e}=d_e^{-1}$. For the realistic mass ratio and $k_{0e}=d_i^{-1}$, early time compressive heating contribution from \ptheta~ dominates \PID~ even at the end of the simulation at $t=6\tau_0$.

\begin{figure}
    \includegraphics{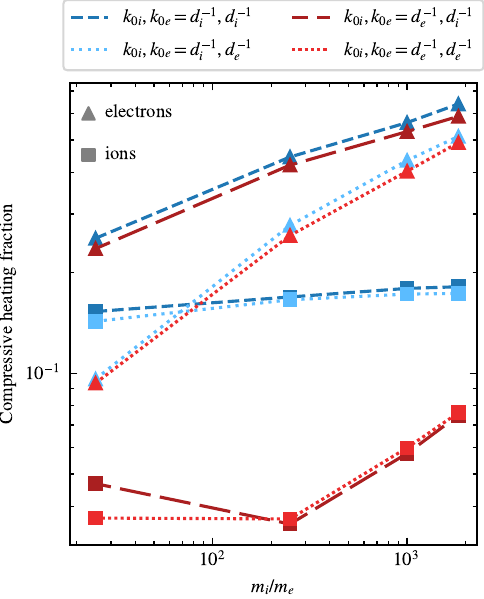}
    \caption{The compressive heating fractions for each species and simulation. Electrons are marked with triangles, and ions are marked with squares. Compressive heating fractions are calculated as $\langle\frac{\int_0^t  p\theta \,dt}{\int_0^t \left(p\theta+\Pi D\right) \, dt}\rangle$ over the interval $4\tau_0\leq t\leq 6 \tau_0$ and represent what proportion of the heating is due to compressive dynamics.}
    \label{fig:compressibility}
\end{figure}

We further quantitatively compare the relative contributions of \PID~ and \ptheta. Figure~\ref{fig:compressibility} shows the average ratio of the two quantities ($\langle{\int_0^t  p\theta \,dt}/{\int_0^t \left(p\theta+\Pi D\right) \, dt}\rangle$) at late times ($4\tau_0\leq t\leq 6 \tau_0$). Triangles represent the values for electrons while squares represent values for ions. The colorscheme is consistent with the previous figures. This figure reinforces the conclusion that varying the $k_0$ of a given species has the most significant effect on its heating properties while varying the other species' $k_0$ does not impact the heating fraction significantly. As the cascade of energy first sees the ions, they potentially heat first and the left-over energy cascades to electron scales. This implies that the ion-closure can indirectly have significant effects on electron heating. This becomes more apparent at smaller mass-ratios and is also supported by Fig. \ref{fig:big_energies}, where we see the ion closure parameter having a notable effect on the electron thermal energy.

The compressive heating fraction for the ions is the smallest for the cases when $k_{0i}=d_e^{-1}$. The cases with $k_{0i}=d_i^{-1}$ have about three times more compressive heating fraction. Interestingly, these are the cases which show the least variation for the ions. The electrons show a significantly more variation with $k_{0e}$ and mass ratio. Between the two possible values of $k_{0e}$, the cases with $d_i^{-1}$ show smaller variation compared to the $d_e^{-1}$ cases. All the cases approach an almost consistent value of compressive heating fraction for electrons for the realistic proton-electron mass ratio, consistent with the previous conclusions \citep{edyvean24scale}.

\begin{figure*}
    \centering
    \includegraphics[width=7in]{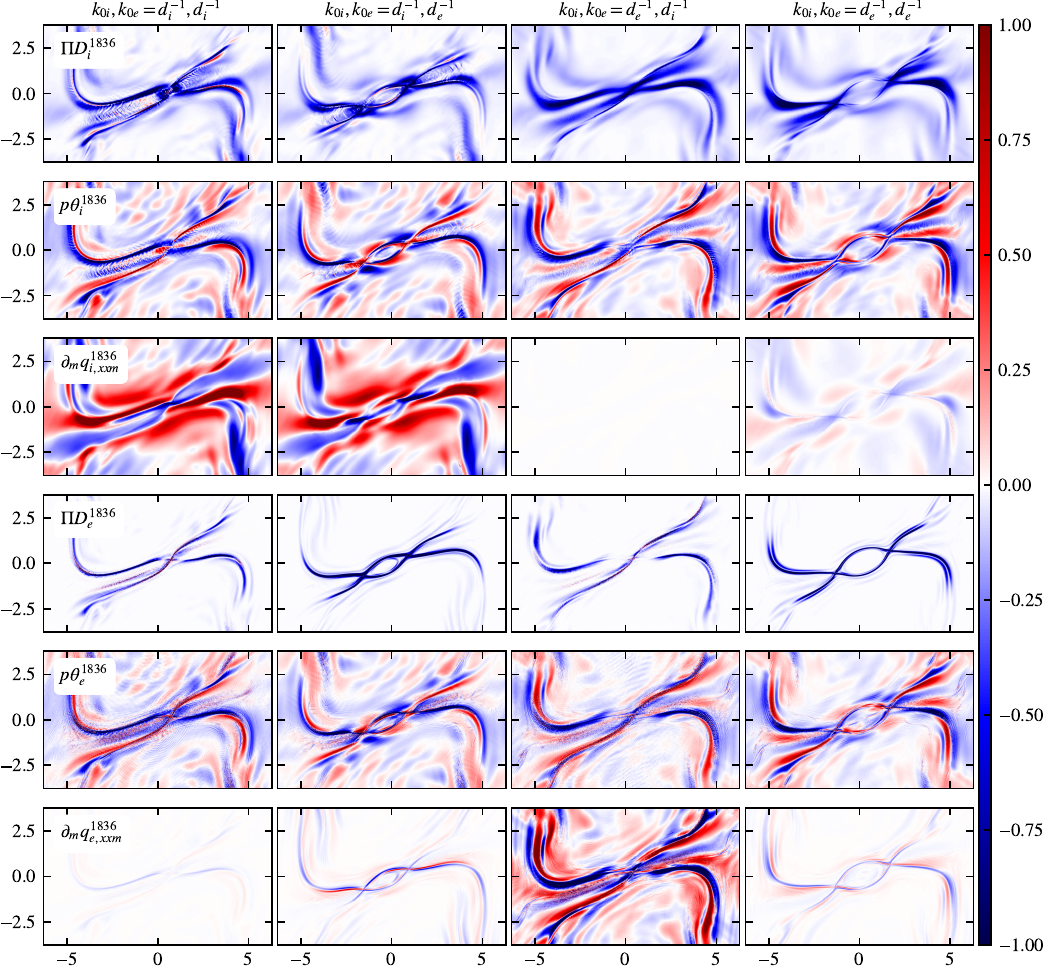} 
    \caption{Snapshots of $\Pi D$, $p\theta$ and  $\partial_m q_{xxm}$ of ions and electrons for different combinations of $k_{0e,0i}$, with $m_i/m_e=1836$. Snapshots were taken at $t=4.9\tau_0$, and zoomed to the region$x\in[-2\pi d_i, 2\pi d_i], y\in[6.8\pi d_i, -6.8\pi d_i]$, with the $y$ axis rolling across the periodic boundary of the simulation. (The values on the y-axis of these plots are the distance from the simulation boundary, $y=\pm4\pi d_i$). A given species variable is normalised by magnitude across all $k_{0i,0e}$ combinations, ignoring the top $1\%$.}
    \label{fig:1836dQpguZOOMED}
\end{figure*}

Figure \ref{fig:1836dQpguZOOMED} shows \PID~, \ptheta~, and \divq, for both ions (top three rows) and electrons (bottom three rows), for  a small portion of the simulation domain near an intense reconnection site that develops at late time ($t=4.9\tau_0$) in the realistic mass ratio simulation. The four columns show four different combinations of $k_{0i},k_{0e}$. Each row is normalised by magnitude across all $k_{0i,0e}$ combinations, ignoring the top $1\%$. Two features stand out: i) the formation of a plasmoid for cases with $k_{0e}=d_e^{-1}$ and, ii) the most intense values  of each variable occur near the separatrices of the reconnection site as well as the edges of the plasmoid.

Generally, the large scale structure is the same across all columns but the small scale structure varies drastically, even after accounting for the plasmoid. The cases with $k_{0i}=d_i^{-1}$ (first two columns) have significantly larger \divqi~ and ``ripple''-like structures in \PID$_i$ and \ptheta$_i$ reminiscent of fully kinetic PIC simulations \citep[see e.g. Fig 6 of][]{yangMNRAS2019}. This feature is significantly suppressed for the cases with $k_{0i}=d_e^{-1}$ (right columns). 

\begin{figure*}[t!]
    \centering
    \includegraphics[width=7in]{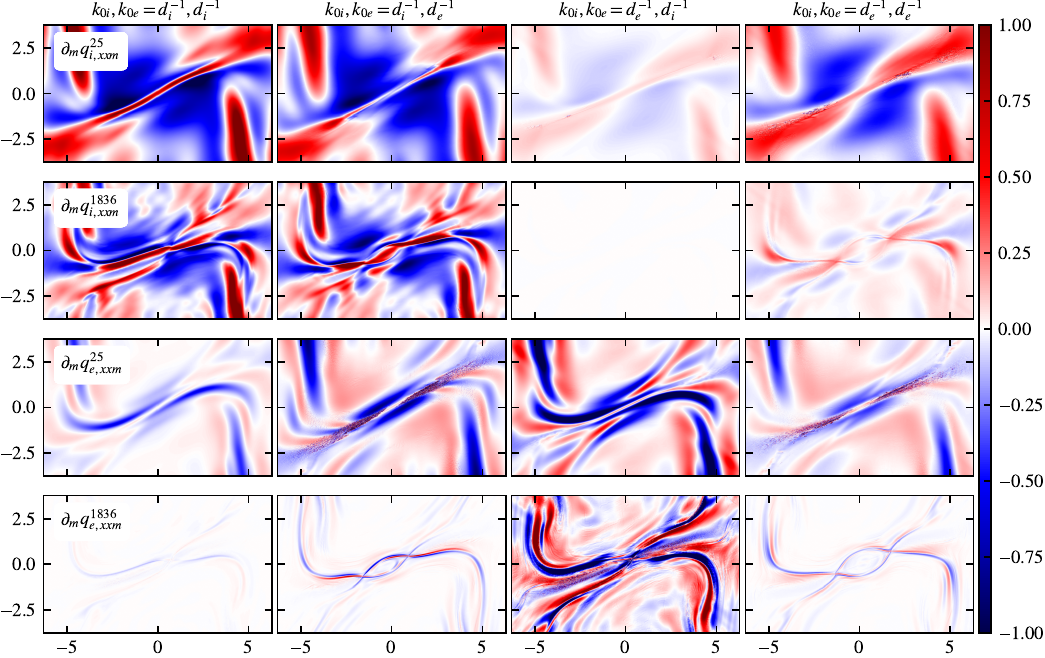}
    \caption{Snapshots of $\partial_m q_{xxm}$ of ions and electrons for different combinations of $k_{0i,0e}$, for $m_i/m_e=25$ and $1836$. Time within simulation, zoom, and normalisation are the same as \ref{fig:1836dQpguZOOMED}}
    \label{fig:MR_comp_qxx_ZOOMED}
\end{figure*}

In Figure \ref{fig:MR_comp_qxx_ZOOMED} we show how $\partial_m q_{\alpha,xxm}$ varies across the same region of our domain as in Figure \ref{fig:1836dQpguZOOMED}, for simulations with $m_i/m_e=25$ and $1836$ for different combinations of $k_{0i,0e}$ at a time $t=4.9\tau_0$. We see the expected increase in intensity and localisation of structure, for both ions and electrons, around the current sheets as we move from $m_i/m_e=25$ to $1836$. The large-scale statistical behaviour is similar for both mass ratios. For all $k_{0i,0e}$ combinations and across both mass ratios, the electron structures are significantly more localised than that of the ions. Both ions and electrons see the biggest effect on the localisation of their divergence of heat flux when varying the electron closure parameter. This is significantly more pronounced for the electrons.

\divqi~ shows a significant change in magnitude when $k_{0i}$ is changed from $d_i^{-1}$ to $d_e^{-1}$ (top two rows of Figure \ref{fig:MR_comp_qxx_ZOOMED}). For electrons, the behaviour is dependent on both $k_{0i}$ and $k_{0e}$. This effect is more pronounced for larger mass ratios. A likely reason for this is that increased mass ratio increases $k_{0e}$ proportional to $\sqrt{m_i/m_e}$, hence affecting the magnitude of \divq~ (see Eqn \ref{equ:divhf}).

\section{Conclusion}\label{Conculsion}
Using high resolution 10-moment simulations varying the mass ratio and the closure parameters for both ions and electrons, we study how the heating of ions and electrons is affected by the choice of closure parameter. \citet{wang2015comparison} showed a good agreement with PIC simulations of reconnection for $k_{0e,0i} = d_e^{-1}$. We expand on this and simulate multiple closure parameter combinations of ions and electrons, see table \ref{tab:comb}. 

We find that the choice of closure parameter impacts the dynamics of collisionless turbulent plasma simulations using a moment hierarchy model. Evidently, the choice of $k_{0\alpha}$ affects $\partial_m q_{\alpha,ijm}$ which in turn affects the pressure tensor. The modifications of the pressure tensor feed into updated pressure-strain interactions, and hence affect the heating of ions and electrons. This ``downward propagation'' of the error into lower-order moments is more significant for smaller mass ratios. 

We find that the ion closure parameter has a greater effect on ion/electron heating. $k_{0i}=d_e^{-1}$ confines the ion heat-flux physics to much smaller scales, especially for larger mass ratio cases. This, however, is not true when $k_{0i}=d_i^{-1}$. The ion heat-flux has significant contributions at ion scales, and the variations in the heat-transfer mechanisms introduced by it affect the electron dynamics, and hence their heating.  Comparisons with fully kinetic simulations are required to conclusively understand the physical processes involved.

These conclusions add a nuance to our findings in the precursor study \cite{edyvean24scale} where we found that a mass ratio of 250 was sufficient to obtain consistent ion/electron heating. As shown in this manuscript, the relative heating rates are sensitive to the closure parameter and do not asymptote to realistic values for smaller mass ratios. 

While we were not able to perform equivalent fully kinetic simulations for the lack of computational resources, a comparison of the different closure parameter combinations will help identify the most relevant closure parameters that mimic the physics captured by fully kinetic simulations. Another possible extension would be to compare updated closure models for the 10-moment framework, such as the one by \citet{ng2020improved}, with fully kinetic simulations.An interesting open question is whether the scale for $\partial_mq_{ijm}$ is the inertial length $d{_i,e}$ or if it is the gyro-radius $\rho_{i,e}$. Further simulations varying the plasma $\beta$ are needed to resolve this issue. 

Another possible investigation to identify the ``downward propagation'' of the errors in the moment hierarchy is to employ Hermite decomposition of the velocity-space which allows one to change the number of moments parametrically via the number of Hermite modes retained in the simulation \citep{delzanno2015JCP, vencels2016JoPCS, roytershteyn2018FASS,koshkarov2021CPC}. The closure in such models is also an open area of research. For example, \textcite{issan2024conservative} show that a direct closure model for symmetrically weighted Hermite methods does not exist that can conserve mass, momentum, and energy simultaneously. However, the effects of the higher moment errors can be minimized by retaining more Hermite moments.  

\section{Author Contributions}

J.V.H.E: Set up and ran the simulations used in this study, wrote and ran the analysis scripts used in the production the figures, wrote the initial draft of the paper, contributed with edits to the paper. 

A.F.I: Design and testing of initial (small-scale) simulation sweep, identification of effect, modification and finalization of figures, editing.

T.N.P: Supervised J.V.H.E and A.F.I, helped refine the problem setup, discussed the results and potential analyses, edited the manuscript. 

\section{Acknowledgements}

J.E, A.F.I, and T.N.P would like to thank the team at the Victoria University of Wellington's R$\Bar{a}$poi computer cluster for providing the capability to run and analyse the simulations used for this study.

\bibliography{main} 

\bibliographystyle{aipnum4-2} 

\end{document}